# Control of conductivity in Fe-rich cobalt-ferrite thin films with perpendicular magnetic anisotropy

Masaya Morishita,[1)] Tomoyuki Ichikawa,[1)] Masaaki A. Tanaka,[1,*)] Motoharu Furuta,[1)] Daisuke Mashimo,[1)] Syuta Honda,[2,3)] Jun Okabayashi,[4)] and Ko Mibu[1)]

[1] *Graduate School of Engineering, Nagoya Institute of Technology, Nagoya, Aichi 466-8555, Japan*

[2] *Department of Pure and Applied Physics, Kansai University, Suita, Osaka 564-8680, Japan*

[3] *Center for Spintronics Research Network, Graduate School of Engineering Science, Osaka University, Toyonaka, Osaka 560-8531, Japan*

[4] *Research Center for Spectrochemistry, The University of Tokyo, Bunkyo-ku, Tokyo 113-0033, Japan*

\* Corresponding author

**Abstract**

We fabricated two types of cobalt-ferrite (001) thin films, insulative Fe-rich cobalt-ferrite $Co_xFe_{3-x}O_{4+\delta}$ (I-CFO) and conductive Fe-rich cobalt-ferrite $Co_yFe_{3-y}O_4$ (C-CFO), with perpendicular magnetic anisotropy (PMA) on MgO (001) substrates. Although the stoichiometric cobalt-ferrite is known as an insulating material, it is found that the conductivity of Fe-rich cobalt-ferrites can be controlled by changing the source materials and deposition conditions in the pulsed laser deposition technique. The I-CFO and C-CFO films exhibit PMA through the in-plane lattice distortion. We investigated the Fe-ion-specific valence states in both I-CFO and C-CFO films by Mössbauer spectroscopy and X-ray magnetic circular dichroism, and found that the difference in conductivity corresponds to the abundance ratio of $Fe^{2+}$ state at the octahedral *B*-site ($O_h$) in the inverse spinel structure. Furthermore, first-principles calculations reproduce the changes in the density of states at the Fermi level depending on the cation vacancies at the *B*-site, which explains the difference in the conductivity between I-CFO and C-CFO.

## I. Introduction

Spinel-type cobalt-ferrites ($CoFe_2O_4$: CFO) have been investigated for a long time for magnetic applications as a magnetic insulator with high-frequency performance, magneto-strictive properties, and so on, in the forms of bulk, powder, and thin films, because of their insulating advantages [1-13]. They have an inverse spinel structure in $AB_2O_4$ chemical formulation, with $Fe^{3+}$ at the tetragonal *A*-site ($T_d$), and $Fe^{3+}$ and $Co^{2+}$ at the octahedral *B*-site ($O_h$). Recent developments in the spintronics boost the application of thin CFO films stacked with other nonmagnetic and magnetic layers, which are, for example, utilized as spin-filtering tunnel barriers [14-19]. There are some reports that CFO (001) films exhibit perpendicular magnetic anisotropy (PMA) under the tensile epitaxial strain [20-35]. The materials with PMA are strongly demanded for the high-density recording techniques. Therefore, CFO films can be one of the best candidates for the PMA system without rare-earth elements. The PMA in the CFO films is thought to originate from $Co^{2+}$ ($3d^7$) at the octahedral *B*-site ($O_h$) in the tetragonally distorted inverse spinel structure [30]. Furthermore, it is reported that the squareness of hysteresis curves, i. e., the abruptness in magnetization reversal, is improved in Fe-rich CFO films [28]. The valence states of the cations in these stoichiometric and Fe-rich CFO films can be examined by site-specific analyses such as X-ray magnetic circular dichroism (XMCD) [25,35] and Mössbauer spectroscopy (MS) [7,33]. For the charge neutrality, the formal valence states are recognized as $Co^{2+}$ ($O_h$), $Fe^{3+}$ ($O_h$), and $Fe^{3+}$ ($T_d$) in stoichiometric CFO films [9]. However, the substitution of $Co^{2+}$ by $Fe^{2+}$ promotes the formation of $Fe^{2+}$ ($O_h$) sites, which induces the conductive properties in CFO films. The conductivity in CFO films would arise depending on the compositions due to the formation of $Fe^{2+}$ states through the $Co^{2+} \rightarrow Fe^{2+}$ substitution, which is regarded as an approach to magnetite $Fe_3O_4$. On the other hand, insulative CFO can be grown by the $Co^{2+} \rightarrow Fe^{3+}_{2/3}\square_{1/3}$ ($\square$ = cation vacancies) substitution, which is an approach to insulative maghemite $\gamma$-$Fe_2O_3$ [4,33,36]. Hereafter, we refer to insulative Fe-rich cobalt-ferrite $Co_xFe_{3-x}O_{4+\delta}$ as "I-CFO" and conductive Fe-rich cobalt-ferrite $Co_yFe_{3-y}O_4$ as "C-CFO". The cation vacancies in I-CFO are expressed as excess oxygen with the symbol of $\delta$.

The I-CFO films with PMA can be used as spin-filtering tunnel barriers to create perpendicularly spin-polarized electronic currents [19]. For this purpose, the epitaxial I-CFO (001) films with in-plane tensile strain have to be grown on conductive underlayers. In order to develop the fully epitaxial spin-filtering magnetic tunnel junctions (MTJs), the interfacial matching between conductive underlayers and I-CFO insulating barriers has to be examined carefully. From the viewpoints of the lattice match and PMA stabilization, the C-CFO layers with PMA are highly promising as conductive electrode layers under the I-CFO barriers. Although the use of $Fe_3O_4$, which is conductive through the electron hopping between $Fe^{2+}$ and $Fe^{3+}$ ions, as electrode layers of conventional MTJs with nonmagnetic barriers are reported recently [37,38], the PMA properties can also be appended to the oxide spintronics researches by using the Fe-rich I-CFO and C-CFO films. In fact, the MTJ combining both I-CFO tunnel barriers and C-CFO electrode layers achieved a large tunnel magnetoresistance ratio of −20% at 100 K [19]. In order to develop I-CFO and C-CFO films with PMA for the high-quality Co-ferrite MTJs, the exact selective growth conditions for both I-CFO and C-CFO films and their detailed electronic states have to be examined precisely.

In spite of the realization of large PMA in CFO films on MgO substrates, it is known that the magnetic properties of epitaxial films of spinel ferrites are strongly affected by the existence of anti-phase boundary (APB). Until now, the APB in $Fe_3O_4$ films grown on the MgO substrates fabricated by various methods was examined [39-43]; the APB broadens the abruptness in magnetization curves (MH curves) with the suppression of saturation magnetization $M_s$ values. The APB in CFO films on the MgO substrates is also discussed [4,25]. Although PMA was indicated by the MH curves in these experiments, the magnitude of the $M_s$ is much smaller than that of bulk $CoFe_2O_4$. Therefore, the existence of APB in the spinel-type lattice structure is inevitable and it would affect the MH curves. These facts have to be considered also for the I-CFO and C-CFO cases.

In this paper, we aim to establish the selective film growth methods between insulative Fe-rich cobalt-ferrite and conductive Fe-rich cobalt-ferrite films with PMA, and discuss the origin of conductive properties from the results of transport measurements and site-specific magnetic spectroscopies of XMCD and MS.

## II. Experimental

Single-crystalline I-CFO ($Co_xFe_{3-x}O_{4+\delta}$) and C-CFO ($Co_yFe_{3-y}O_4$) films of 20 nm in thickness with various Co compositions of $x$ ($x$ = 0.00, 0.11, 0.23, 0.36, 0.43, 0.56, 0.66, 0.76, 0.87, and 1.00) and $y$ ($y$ = 0.00, 0.23, 0.42, and 0.66)

were grown on the MgO (001) substrates by a pulsed laser deposition (PLD) method using the source materials with different powder compositions. The source materials for the I-CFO, except $x = 0$, were prepared by mixing $CoFe_2O_4$ and $Fe_3O_4$ powder, and those for the C-CFO, except $y = 0$, were by mixing $CoFe_2O_4$ and α-Fe powder. Pure $Fe_3O_4$ powder was used for the source materials of the I-CFO with $x = 0$ and C-CFO with $y = 0$. The compositions between the Co and Fe atoms were controlled by the ratios of the source materials and confirmed by an electron probe microanalyzer. After the pressing process under 20 MPa, the I-CFO and C-CFO sources were baked in an atmospheric environment for 12 hours at the maximum temperature of 1100°C and 350°C, respectively. Note that the source materials were not perfectly consummated as the aiming I-CFO and C-CFO films because the amounts of oxygen and cation vacancies are expected to be controlled by the film growth conditions.

The PLD was performed using frequency-doubled Nd:YAG laser with a pulse width of 6 ns and a repetition rate of 30 Hz. The energy density of the laser beam was controlled to 1 J/cm$^2$ by an optical lens. The I-CFO films were grown in $O_2$ pressure of 6 Pa at the substrate temperature of 300°C. The C-CFO films were grown in Ar pressure of 4 Pa at the substrate temperature of 300°C. The gas pressure conditions and substrate temperatures during the deposition were optimized to obtain CFO films having large in-plane tensile strain in the (001) crystal orientation, which induces the PMA. The eclipse PLD method, where a shadow mask was placed between the source material and the substrates, was used in order to reduce the formation of droplets and particulates [44].

The crystal structures of the prepared films were investigated by X-ray diffraction (XRD) using a Cu-Kα source. In order to investigate the in-plane and out-of-plane strains, reciprocal space maps of X-ray diffraction intensity were measured for both I-CFO and C-CFO films. The transport properties of prepared films were investigated using the four-probe methods with cooling by a helium cryostat. The MH curves were measured using a superconducting quantum interference device (SQUID) magnetometer. $^{57}$Fe Mössbauer spectroscopy using a radioactive $^{57}$Co source was conducted at room temperature (RT) by the conversion electron detection method. The γ-rays were irradiated from the film normal direction and the emitted electrons were detected by a proportional gas counter. X-ray absorption spectroscopy (XAS) and XMCD were performed at BL-7A in the Photon Factory at the High Energy Accelerator Research Organization (KEK-PF). For the XAS and XMCD measurements, the photon helicity was fixed, and a magnetic field of ±1.2 T was applied parallel along the incident polarized soft X-ray beam, to obtain signals defined as *µ+* and *µ−* spectra. The total electron yield (TEY) mode was adopted, and all the measurements were performed at RT. Since the TEY measurements are surface sensitive which detects the signals beneath 3 nm from the sample surface, we prepared the samples for XAS and XMCD by capping the surfaces by 1-nm-thick Pd layers deposited at RT.

### III. Results
#### A. X-ray diffraction

Figure 1(a) shows the XRD patterns of the typical I-CFO and C-CFO films with various Co composition *x* and *y* ($x$ = 0.00, 0.23, 0.36, 0.66 and $y$ = 0.00, 0.23, 0.42, 0.66), with the scattering vector perpendicular to the film plane. They exhibit clear I-CFO (008) and C-CFO (008) diffraction peaks. This means that the I-CFO and C-CFO films are highly oriented along the [001] directions on the MgO (001) substrates. The I-CFO (008) peaks are shifted to lower angles systematically with increasing *x*, indicating the extension of the lattice along the *c*-axis. On the other hand, the C-CFO (008) peaks are shifted to higher angles systematically with increasing *y*, indicating the suppression of lattice along the

*c*-axis. The additional satellite peaks, which originate from the interference effect, are detected beside the (008) peaks showing high crystalline qualities of the films prepared by the PLD with a shadow mask. Figure 1(b) shows the Co composition dependence of the vertical lattice constant $a_\perp$ of the I-CFO and C-CFO films estimated from the XRD patterns. The red and blue broken lines represent the linear relations in Begard law between bulk $Fe_3O_4$ and $CoFe_2O_4$, and that between bulk $\gamma$-$Fe_2O_3$ and $CoFe_2O_4$, respectively. The out-of-plane lattice constants of the I-CFO films decrease with the decrease of the Co composition $x$, while those of the C-CFO films increase with the decrease of the Co composition $y$. Figure 1(c) shows the reciprocal space maps around the MgO (113) peak in the horizontal Q (110) and vertical Q (001) directions for the I-CFO ($x$ = 0.66) and C-CFO ($y$ = 0.66) films. The I-CFO (226) and C-CFO (226) peaks were observed near the MgO (113) peak. The in-plane Q values of the I-CFO (226) and C-CFO (226) peaks coincide with the value of the MgO (113) peak, which means that the in-plane lattice constants of the I-CFO and C-CFO films are twice that of the MgO substrates (8.42 Å). Note that the tendency of the in-plane lattice constant, which is less dependent on the Co composition, is consistent with previously reported results [19, 35]. From the results of the in-plane $a_{||}$ and out-of-plane $a_\perp$ lattice constants, it is suggested that the I-CFO and C-CFO films have in-plane tensile strain and that the structures are tetragonally distorted to settle the lattice mismatch with the MgO substrates.

### B. Resistivity measurements

Temperature dependence of resistivity for the C-CFO films grown on MgO (001) substrates with different Co composition $y$ is shown in Fig. 2(a). With increasing $y$, the resistivity of the C-CFO films systematically increases. For the I-CFO films, the resistivity was over $10^5$ Ωcm at 300 K for all the Co composition $x$ and they can be regarded as insulators. The resistivity of the C-CFO ($y$ = 0.00), $Fe_3O_4$, film at 300 K is 0.03 Ωcm, which is in the same order as in the previous studies [36, 45]. In order to confirm the hopping conduction through the thermal activation, temperature $T$ dependence of the resistivity $\rho$ is plotted by the Arrhenius equation $\rho = \rho_0 \exp(\Delta E/k_B T)$ and fitted with the least-squares method, as shown in Figs. 2(b)-(e). Here, $\rho_0$ is the pre-exponential factor of the resistivity, $\Delta E$ is the activation energy, and $k_B$ is Boltzmann constant. Except the low-temperature parts for $y$ = 0.00 and 0.23, the resistivity follows the Arrhenius plots, which means that the conduction mechanism of C-CFO is due to the electron hopping between the *B*-site Fe ions with different valences. The deviation of the experimental data from the fitting line for $y$ = 0.00 and 0.23 originates from the Verwey transition. The Verwey transition is a dramatic change in the electrical transport properties accompanied with a structural transition from the cubic phase to the monoclinic phase around 120 K, which suggests that a high-quality $Fe_3O_4$ film is prepared [46,47]. Figures 2(f) and 2(g) exhibit the composition $y$ dependence of $\rho_0$ and $\Delta E$, respectively. The $\Delta E$ corresponds to the activation energy of electron hopping between $Fe^{2+}$ and $Fe^{3+}$ ions. The increase in resistivity with increasing Co in Fig. 2(a) is found to be due to the increase in activation energy of the electron hopping.

### C. Magnetic properties

Figures 3(a)-(d) show the results on magnetization measurements by a SQUID magnetometer with the magnetic field applied along the in-plane and out-of-plane directions at 300 K for the I-CFO ($x$ = 0.00, 0.43, and 0.87) and C-CFO ($y$ = 0.23, 0.42, and 0.66) films. The PMA characteristics are obtained for all cases of the I-CFO and C-CFO films except $x$ =

0.00 ($\gamma$-Fe$_2$O$_3$) and $y$ = 0.00 (Fe$_3$O$_4$). Figure 3(e) shows the Co composition $x$, $y$ dependence of squareness ratio $M_r/M_s$ of I-CFO [33] and C-CFO in the out-of-plane MH curves. The squareness ratio in I-CFO is larger than that in C-CFO for entire Co composition ranges. In I-CFO films, the squareness ratio decreases with the Co composition $x$. For C-CFO films, on the other hand, the squareness ratio increases with Co composition $y$. The difference in tendency on the squareness ratio as a function of Co composition can be explained by the strain magnitude as a function of Co composition. As shown in Fig. 1(b), the out-of-plane lattice constants of I-CFO films increase and those of I-CFO films decrease with the increase of Co composition. The in-plane lattice constants of both I-CFO and C-CFO films almost remain unchanged for all the Co compositions. Therefore, the distortion of I-CFO films decreases, and that of C-CFO films increases with the Co composition $x$. The PMA energies estimated from the area surrounded by the loops of out-of-plane and in-plane loops for I-CFO ($x$ = 0.66) and C-CFO ($y$ = 0.66) are 1.0 and 0.3 MJ/m$^3$, respectively. The deterioration of the squareness in the MH curves and the changes of $M_s$ imply the existence of APB in the films.

### D. Mössbauer spectroscopy

Figure 4 shows the $^{57}$Fe Mössbauer spectra for the I-CFO ($x$ = 0.23) and C-CFO ($y$ = 0.23) films taken at RT using the conversion electron Mössbauer spectrometry (CEMS) mode. Both cases clearly show magnetic spectra, with no component from $\alpha$-Fe or $\alpha$-Fe$_2$O$_3$. In the case of I-CFO, a six-line spectrum pattern with symmetric line heights is observed, which suggests that the film is composed only of the Fe$^{3+}$ ions without Fe$^{2+}$ ions. This is consistent with the resistivity measurement because the hopping conduction is suppressed without Fe$^{2+}$ [48-50]. On the other hand, the spectrum for C-CFO represents an asymmetric and rather broad line shape due to the overlapping of two magnetic components. These spectra are fitted using the Mössbauer parameters of isomer shift ($\delta$), quadrupole splitting ($2\varepsilon$) and hyperfine field ($B_{hf}$). The fitted parameters are listed in Table 1. For C-CFO, two kinds of magnetic components corresponding to the $A$ and $B$-sites in the spinel structure are necessary for the fitting. The $B$-site components are detected as an Fe$^{2.5+}$ sextet because the observation time in Mössbauer spectroscopy is longer than the time scale of electron hopping between Fe$^{3+}$ and Fe$^{2+}$ sites. The combination of two magnetic components with different isomer shift and hyperfine fields clearly reproduce the asymmetric spectra. The obtained Mössbauer parameters are similar to those in the previous reports on Mössbauer spectra of CoFe$_2$O$_4$ [2,7,10,12,51,52] and Fe$_3$O$_4$ [53,54]. From these site-specific analyses, the Fe$^{2+}$ components are thought to be essential for the conductivity in C-CFO.

The intensity ratios of the six lines in both spectra are between 3:0:1:1:0:3 and 3:2:1:1:2:3, which means that the films have perpendicular magnetic anisotropy but that the direction of magnetic moment is not perfectly perpendicular to the film plane at zero external field. This is consistent with the magnetization measurements, where the squareness ratios do not reach 1. The intensity ratio of the 2$^{nd}$ and 5$^{th}$ peak is smaller for the I-CFO film, suggesting that the perpendicular anisotropy is stronger in the I-CFO system.

### E. XAS and XMCD

Figure 5 shows the XAS and XMCD of C-CFO for Fe and Co $L$-edges with different Co compositions. Spectra are normalized by an incident white line beam intensity before the absorption at the samples. The Fe and Co intensities are plotted in the same vertical scales because XAS roughly provides the Fe and Co compositions. Because of the composition ratio of Co:Fe = 1:12 in $y$ = 0.23, the XAS intensities of Co in this film are suppressed. With increasing the Co compositions, the Co $L$-edge intensities are enhanced systematically. XAS and XMCD line shapes for the Fe $L$-edges show distinctive features with differential line shapes due to the three kinds of Fe states ($Fe^{3+}$ in $O_h$, $Fe^{3+}$ in $T_d$, and $Fe^{2+}$ in $O_h$). For the Fe $L$-edges, although the difference in XAS is small, clear differential XMCD line shapes are detected. The $Fe^{3+}$ state with $T_d$ symmetry exhibits the opposite XMCD sign, which is common for the spinel ferrite compounds. The $Fe^{2+}$ component at 708.0 eV decreases in high Co compositions. In the case of $Fe_3O_4$, the $Fe^{2+}$ component is more enhanced [38]. Therefore, the conductive properties are related to the amounts of $Fe^{2+}$ states, whose quantitative analysis is performed by using the ligand-field-multiplet (LFM) calculation in later. The cases of I-CFO are also shown in Fig. 6, which is similar to the previous reports [35]. Large XMCD signals in Co $L$-edge in spite of small XAS intensity correspond to the saturated magnetized states. Within the orbital sum rule, the large orbital magnetic moments ($m_{orb}$) come from the asymmetric XMCD line shapes. Since the Co site is almost identified as $Co^{2+}$ ($O_h$) symmetry, the sum rules can be applicable for the Co XMCD spectra. We note that the sum rule analysis for the entire Fe $L$-edge XMCD cannot be applicable because it includes three kinds of components, which has to be deconvoluted into each component because XAS/XMCD originates from the atomic excitation process [55]. The spin and orbital magnetic moments for $Co^{2+}$ sites in $y = 0.23$ sample are estimated as 1.32 and 0.63 $\mu_B$, respectively, using an electron number of 7.1 [33]. Large $m_{orb}$ originates from the orbital degeneracy in the $d^7$ electron system in strained Co sites and contributes to the PMA [56]. The strained Co sites are also confirmed by the extended x-ray absorption fine structures [35].

For the analysis of XMCD spectra, we employed the LFM cluster-model calculations including the configuration interaction for the Fe sites in $Co_yFe_{3-y}O_4$ of $y = 0.23$ and 0.66 as tetrahedral ($T_d$) $TMO_4$ and octahedral ($O_h$) $TMO_6$ clusters, modeled as a fragment of the spinel-type structures. The line shapes of $Fe^{2+}$ ($O_h$), $Fe^{3+}$ ($O_h$) and $Fe^{3+}$ ($T_d$) are calculated in the previous reports in the I-CFO using the Coulomb interaction $U$ of 6.0 eV [35] by considering the chemical shift of energy position through the electron number. Here, we adopt this analysis for the estimations of the ratio in $Fe^{2+}$ and $Fe^{3+}$ intensities.

As shown in Fig. 7, the spectral line shapes of XMCD can be reproduced by the LFM calculations qualitatively, at least the peak positions, with three kinds of Fe states. The XMCD spectra for different valences and symmetries are calculated by using the same parameters as the cases of I-CFO [35]. The peak position of $Fe^{2+}$ (709.0 eV) is composed of not only $Fe^{2+}$ but also $Fe^{3+}$ ($O_h$) contribution. We fitted the spectra using these LFM-calculated spectra by changing the ratios of three components. Then, the ratios of $Fe^{2+}$ ($O_h$), $Fe^{3+}$ ($O_h$) and $Fe^{3+}$ ($T_d$) can be estimated to be 1:1:1 for $y = 0.23$, and 0.2:1:1 for $y = 0.66$. The fitting results indicate that the intensity ratio of $Fe^{3+}$ ($O_h$) and $Fe^{3+}$ ($T_d$) is not affected by the Co ion substitution. Therefore, the site-specific analysis in XMCD clearly deduces the contributions of $Fe^{2+}$ states, although it might be overestimated, which play an essential role for the conduction mechanism in C-CFO films.

## F. DFT calculation

In order to investigate the conductive properties of $Co_yFe_{3-y}O_4$ films from the viewpoint of the electronic structures, we performed the first-principles density-functional-theory (DFT) calculations with periodic boundary conditions for an optimized $Co_1Fe_{11}O_{16}$ super-cell structure shown in Fig. 8(a), which was considered as the C-CFO of $Co_{0.25}Fe_{2.75}O_4$ ($y = 0.25$). The lattice constant for the unit cell of 8.35 □ was employed which includes the strain estimated from the XRD experiments. The unit cell includes the $A$- and $B$-sites of 4 and 8 cations, respectively, and these correspond to 4 $Fe^{3+}$ ions at the $A$-sites, and 4 $Fe^{3+}$ ions, 3 $Fe^{2+}$ ions, and 1 $Co^{2+}$ ions at the $B$-sites. The DFT calculations were performed using the VASP code with the projector augmented wave (PAW) potentials [57] in the generalized gradient approximation (GGA) - perdew burke ernzerhof (PBE) [58,59] including the Coulomb repulsion energy $U$ of 6.0 eV. The cutoff energy is set to 400 eV and the crystal momentum $k$ sampling mesh is set to (21, 21, 17).

Figure 8(b) shows the calculated spin-dependent density of states (DOS) for $Co_1Fe_{11}O_{16}$ as C-CFO. The spin-down states of the $B$-site in Fe cross the Fermi level ($E_F$) and contribute to the conductivity. The partial DOS of Co site is located at deeper level and does not contribute to the formation of the intensity at the $E_F$. The band gap between occupied and unoccupied spin-up states is estimated to be 2.58 eV. The DOS clearly indicates the half-metallic property. The Fe magnetic moment of −4.15 $\mu_B$ for $A$-sites and 4.02 $\mu_B$ for $B$-sites can be estimated. The local magnetic moment of 2.70 $\mu_B$ of Co site is also estimated, which is smaller than those of Fe sites and qualitatively consistent with that of the XMCD analysis.

We also calculated the spin-dependent DOS for an optimized $Co_1Fe_{10}O_{16}$ super-cell structure as shown in Fig. 8(c), which is considered as the I-CFO of $Co_{0.25}Fe_{2.50}O_4$ or $Co_{0.273}Fe_{2.727}O_{4.364}$ ($x = 0.273$, $\delta = 0.364$). The calculated DOS is shown in Fig. 8(d) and the insulating property can be seen. The band gaps between occupied and unoccupied states in spin-up and spin-down cases are estimated to be 2.67 eV and 2.40 eV, respectively, whose entire DOS is similar to the previous results of I-CFO [33]. Note that this spin-dependent difference in the band gaps can be applied for the generation of the spin-polarized electrons. In fact, we reported that electrons tunneling through the I-CFO thin films were spin-polarized because of the difference in tunnel probability of spin-up and spin-down electrons caused by the difference in the band gaps [19]. The half-metallic property is guaranteed in the C-CFO, as in the case of $Fe_3O_4$, where $B$-site DOS contribute the conduction at the $E_F$ [38].

## IV. Discussion

The conduction mechanism, structural deformation, and advantages in I-CFO and C-CFO can be discussed using the above results. *First*, MS and XMCD reveal the existence of $Fe^{2+}$ states in C-CFO, which is responsible for the conduction as analogous to the case of $Fe_3O_4$. However, for I-CFO, small amounts of $Fe^{2+}$ states are still detected in XMCD, which is not detected in MS as shown in Fig. 4. It may come from the difference in probing regions; while the MS detects the bulk information because of the larger escape depth of the conversion electrons (~ 100 nm) in comparison with the film thickness (20 nm), XMCD in soft X-ray detects the surface regions beneath 3 nm from the sample surface. Since the valence states of surface regions are slightly modulated and the surface oxygen reduction might occur. However, it is not significant but an offset factor for the detailed analysis of $Fe^{2+}$ states to estimate the chemical compositions. Therefore, site-specific magnetic spectroscopies by MS and XMCD provide the complementary information for the analyses of I-CFO and C-CFO.

*Second*, the conduction mechanism can be discussed by combining the resistivity and DFT calculation. As an analogous system with $Fe_3O_4$, the conduction in C-CFO is caused by the electron hopping between $Fe^{2+}$ and $Fe^{3+}$ in the $B$-site through the activation type excitation. Temperature dependence of the resistivity in Fig. 2 is different from the small polaron formation and variable range hopping schemes [48]. The double-exchange mechanism can be anticipated in the $t_{2g}$ states. Half-metallic nature in the DFT calculations can also be explained by the electron hopping mechanism, which is different from the noble metals and distinctive for conductive oxide materials such as perovskite Mn oxide compounds [60]. Yasui *et al*. reported that a large tunnel magnetoresistance (TMR) effect was obtained for $Fe_3O_4$/MgO/Fe MTJs and that the large

TMR effect derives from the half-metallic nature of $Fe_3O_4$ [38]. Thus the half-metallic C-CFO films may become a good candidate for the spin injector in spintronics devices.

*Finally*, we discuss the selective growth of two types of Fe-rich CFO, C-CFO and I-CFO. The element- and site-selective XMCD and MS clearly suggest the valance states of $Fe^{2+}$ is an essential factor for controlling the conductivity of CFO. For the growth of Fe-rich CFO films, the valance states of Fe ions can be controlled by changing the oxygen flow rate, the source materials, and the substrate temperature during the growth with the deposition atmosphere. The depositions under the $O_2$ pressure with high growth temperature promote the oxidation of $Fe^{2+}$ states. On the other hand, the depositions under the Ar pressure with low growth temperature suppress the oxidation of $Fe^{2+}$ states. In the case of C-CFO, the XMCD results (Fig.7) show that the abundance ratio of $Fe^{2+}$ ions in the *B*-site decreases with increasing the Co composition *y*, which suggests that the $Co^{2+}$ ions in $CoFe_2O_4$ are replaced with $Fe^{2+}$ ions. The XRD results (Fig. 2(b)) also display that the vertical lattice constant of C-CFO films increases with decreasing *y*. From these results, the C-CFO can be recognized as a magnetite-type CFO because their bond distances are tuned between $CoFe_2O_4$ and $Fe_3O_4$ according to the *y* values, following qualitatively to the Begard law. On the other hand, in the case of I-CFO, the XRD result shows that the vertical lattice constant of I-CFO films decreases with decreasing *y* [33]. Therefore, I-CFO can be considered as a maghemite-type CFO because of the tuning of $CoFe_2O_4$ and $\gamma$-$Fe_2O_3$ compositions according to *x* values. The junction composed of I-CFO tunnel barriers and C-CFO electrode layers can be applicable for the spintronics researches and demonstrated for the spin-filtering effect with PMA [19].

## V.     Summary

We fabricated two types of Fe-rich cobalt-ferrite thin films, insulative $Co_xFe_{3-x}O_{4+\delta}$ (I-CFO) and conductive $Co_yFe_{3-y}O_4$ (C-CFO), with PMA. Although the stoichiometric cobalt-ferrites are known as an insulating material, it is found that the conductivity in Fe-rich CFO can be controlled by changing the source materials and deposition conditions in the PLD technique. The I-CFO and C-CFO films also exhibit the PMA through the in-plane lattice distortion. We investigated the Fe-ions-specific valence states in I-CFO and C-CFO films by Mössbauer spectroscopy and XMCD, and found that the difference in conductivity corresponds to the abundance ratio of $Fe^{2+}$ state at the octahedral ($O_h$) site. Furthermore, first-principles band-structure calculation suggested that the electronic structures of C-CFO are half-metallic characteristics and reproduced the difference in the DOS depending on the cation vacancies at the *B*-site in inverse spinel structures, which explains the difference in the conductivity between I-CFO and C-CFO. These add novel functionalities in the oxide spintronics using Fe-rich CFO films.

## Acknowledgments

This work was partly supported by The Telecommunications Advancement Foundation, Kato Science foundation, Yamada Science foundation, and JSPS KAKENHI (No. 22H04966). Parts of the synchrotron radiation experiments were performed under the approval of the Photon Factory Program Advisory Committee, KEK (No. 2021G069).

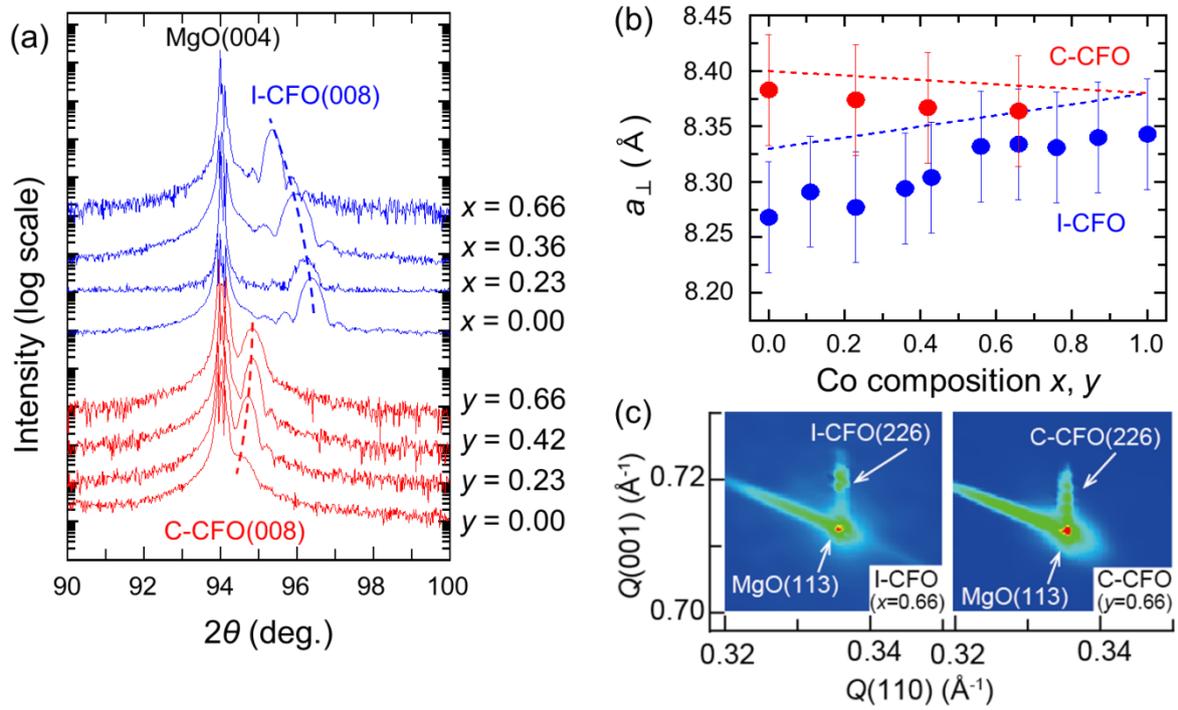

Fig. 1. (a) XRD patterns around the CFO(008) for the $Co_xFe_{3-x}O_{4+\delta}$ (I-CFO) and $Co_yFe_{3-y}O_4$ (C-CFO) films with the scattering vector perpendicular to the film plane. (b) Vertical lattice constants $\boldsymbol{a}_\perp$ of the I-CFO and C-CFO films as functions of the Co composition $x$ and $y$. Red and blue broken lines represent the linear relations in Begard law between bulk $Fe_3O_4$ and $CoFe_2O_4$, and that between bulk $\gamma$-$Fe_2O_3$ and $CoFe_2O_4$, respectively. (c) Reciprocal space maps around the MgO(113) peaks for the I-CFO ($x$ = 0.66) and C-CFO ($y$ = 0.66) films.

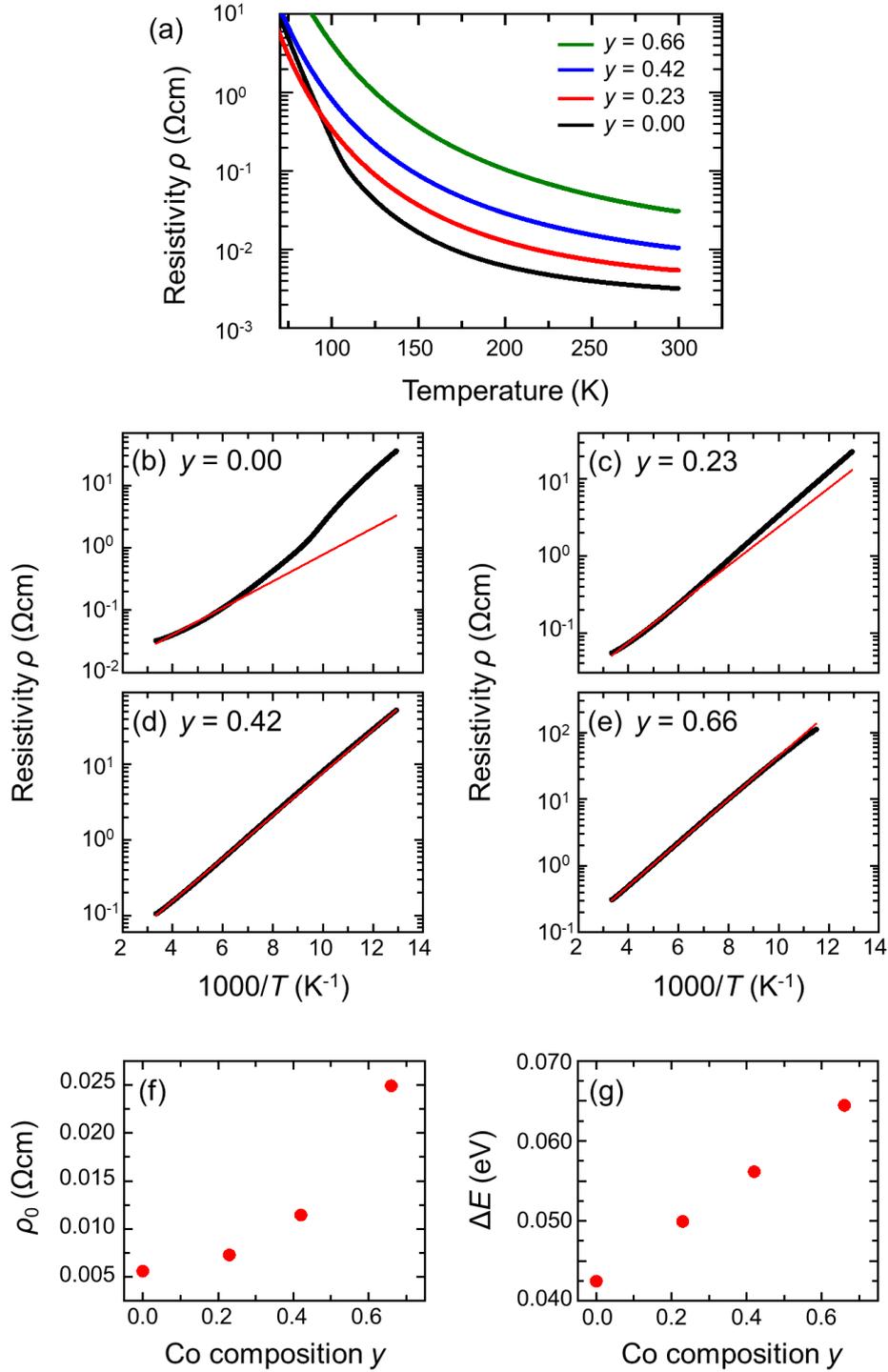

Fig. 2. (a) Electrical resistivity as a function of temperature for conductive $Co_yFe_{3-y}O_4$ films, (b-e) Plot of $\ln\rho$ versus $1/T$ with a linear fitting, (f, g) Pre-exponential factor of the resistivity $\rho_0$ and activation energy $\Delta E$ as a function of Co composition $y$.

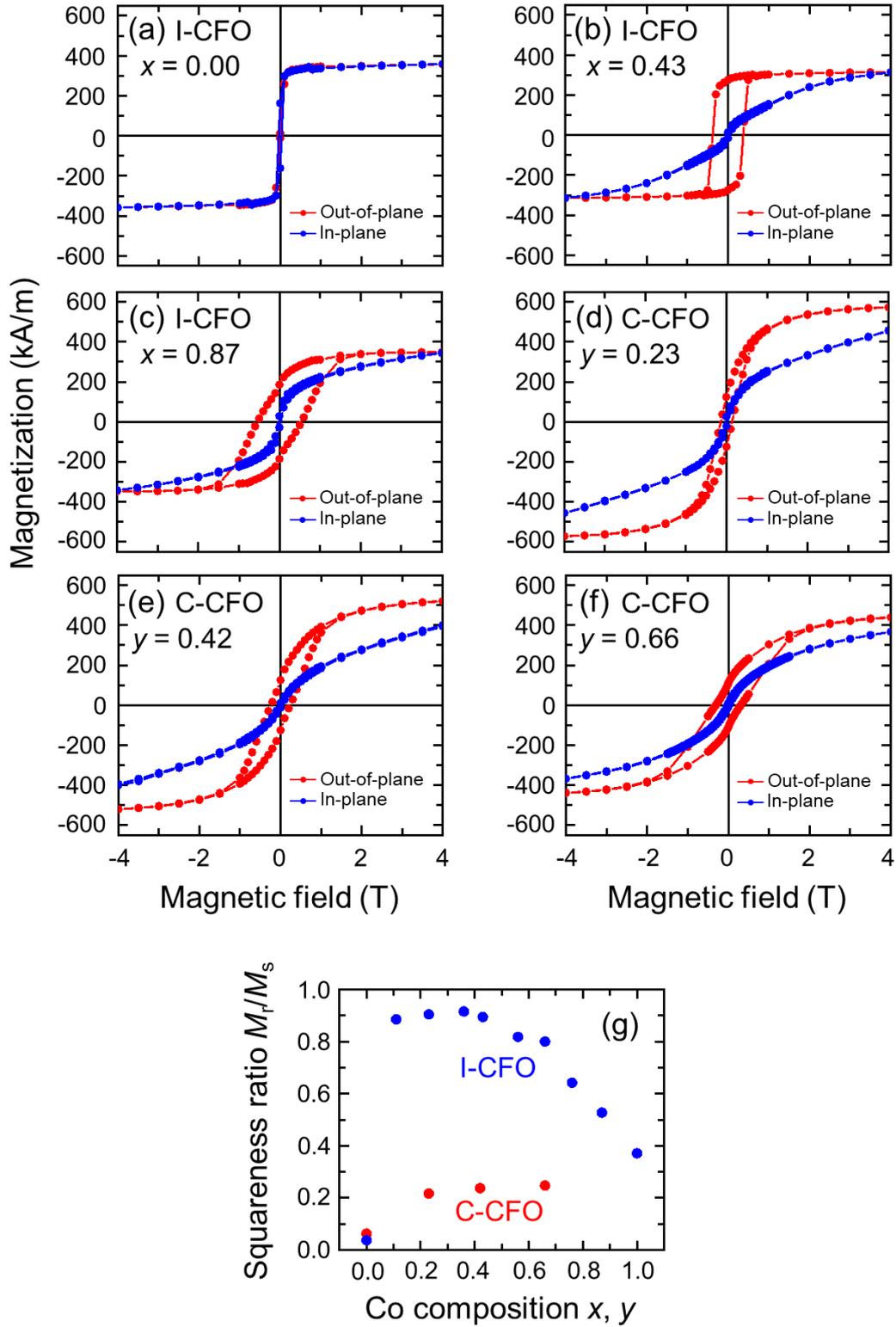

Fig. 3. Out-of-plane and in-plane MH curves of the $Co_xFe_{3-x}O_{4+\delta}$ ((a) $x = 0.00$, (b) $x = 0.43$, and (c) $x = 0.87$) and $Co_yFe_{3-y}O_4$ ((d) $y = 0.23$, (e) $y = 0.42$, and (f) $y = 0.66$) films at 300 K. (g) Squareness ratio of remanent and saturation magnetizations $M_r/M_s$ as a function of Co composition for I-CFO and C-CFO films.

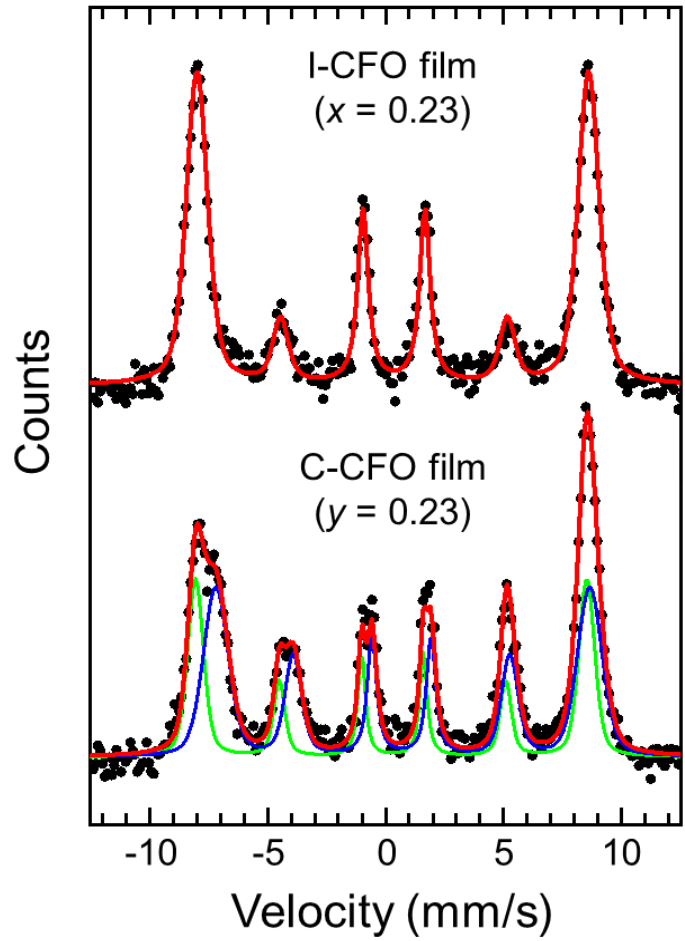

Fig. 4. Mössbauer spectra for insulative $Co_{0.23}Fe_{2.77}O_{4+\delta}$ (I-CFO) and conductive $Co_{0.23}Fe_{2.77}O_4$ (C-CFO) films taken at RT. Dots represent the experimental data. Lines are fitting results. For C-CFO, two components are used and sum of these are also shown.

Table 1 Mössbauer fitting parameters for I-CFO and C-CFO films. Isomer shifts ($\delta$) were defined with respect to α-Fe at RT.

| Sample | Subspectra | $\delta$ (mm/s) | $2\varepsilon$ (mm/s) | $B_{hf}$ (T) | Area (%) |
|---|---|---|---|---|---|
| I-CFO | $Fe^{3+}$ | 0.33 | -0.06 | 51.4 | —— |
| C-CFO | $Fe^{3+}$ A site | 0.26 | -0.06 | 51.4 | 39.8 |
|  | $Fe^{2.5+}$ B site | 0.70 | 0.05 | 49.1 | 60.2 |

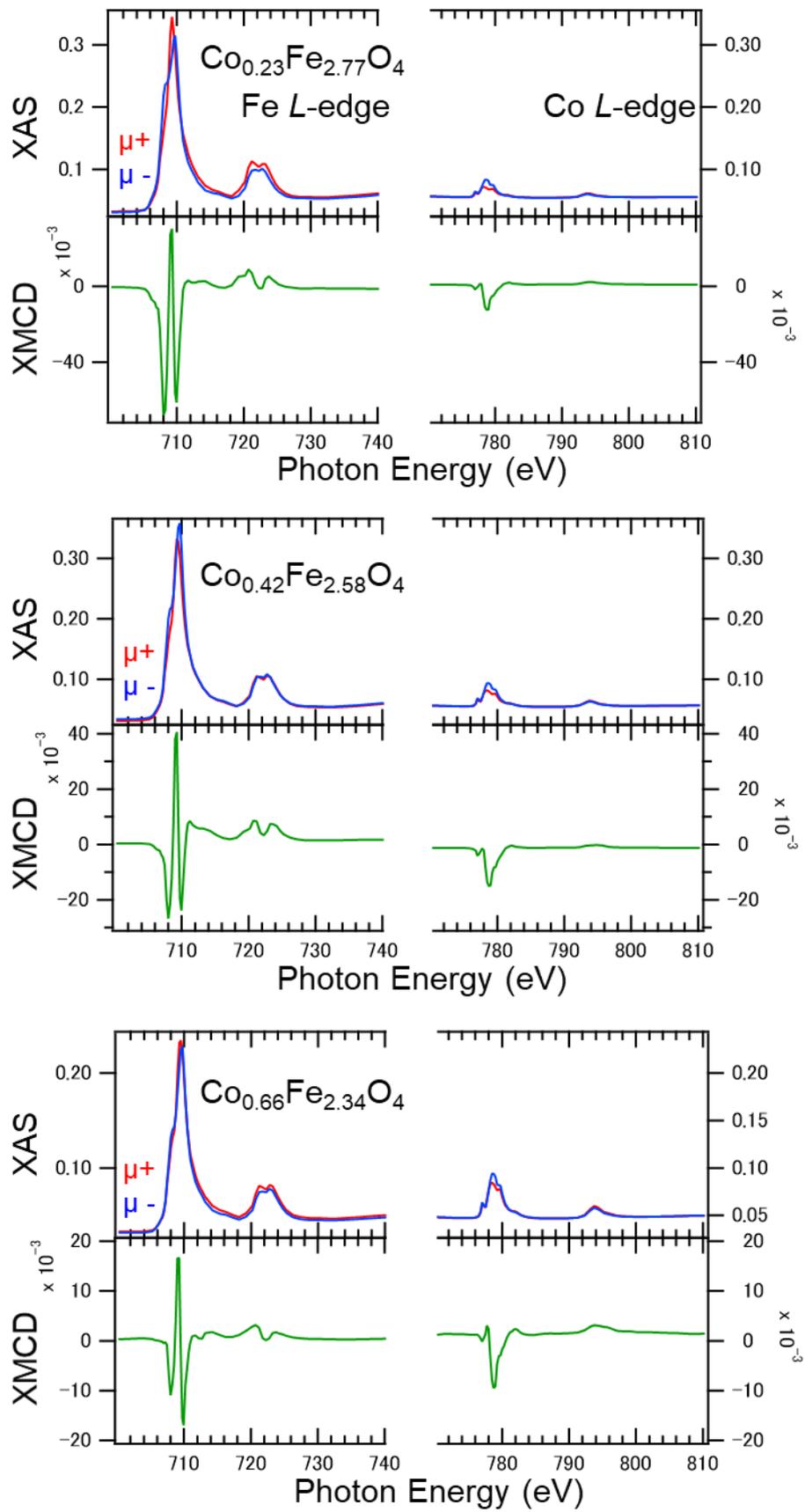

Fig. 5 XAS and XMCD of conductive $Co_yFe_{3-y}O_4$ films for $y$ = 0.23, 0.42, and 0.66.

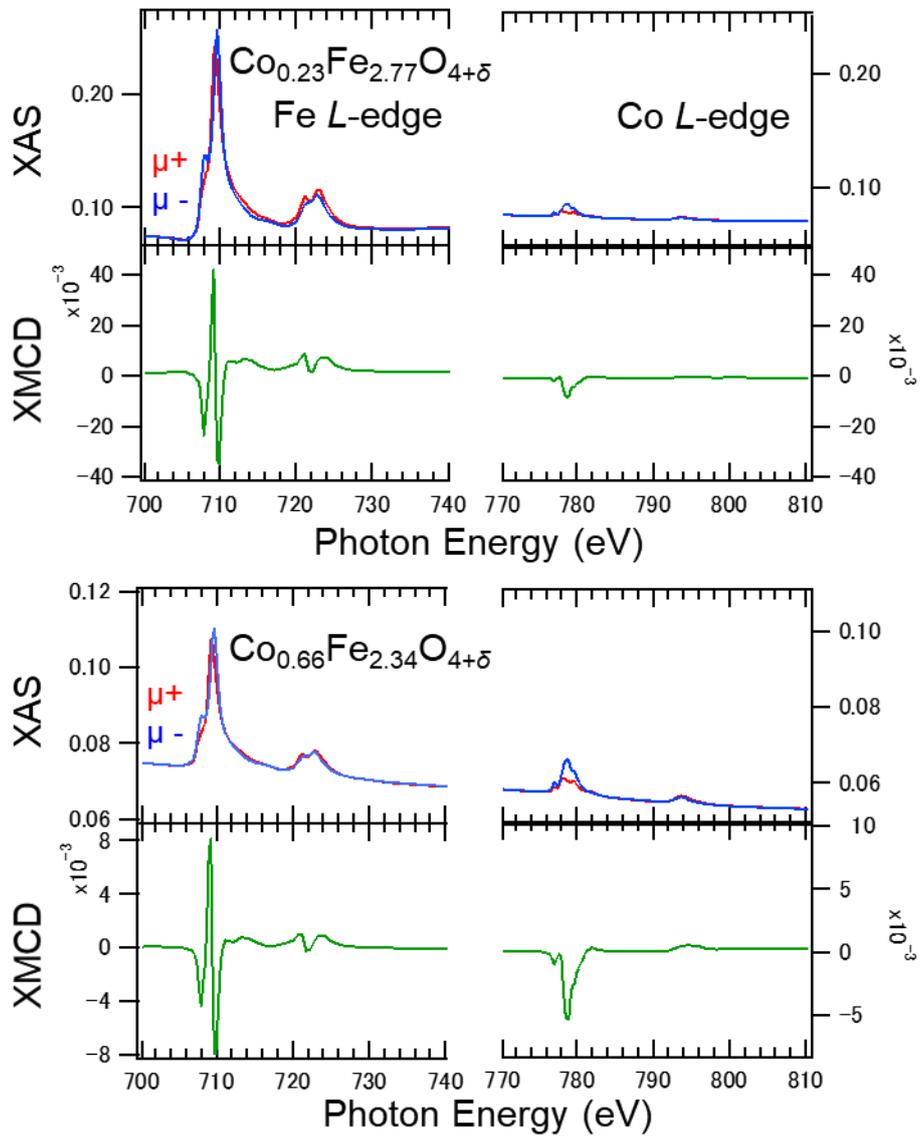

Fig. 6 XAS and XMCD of insulative $Co_xFe_{3-x}O_{4+\delta}$ films for $x$ = 0.23 and 0.66.

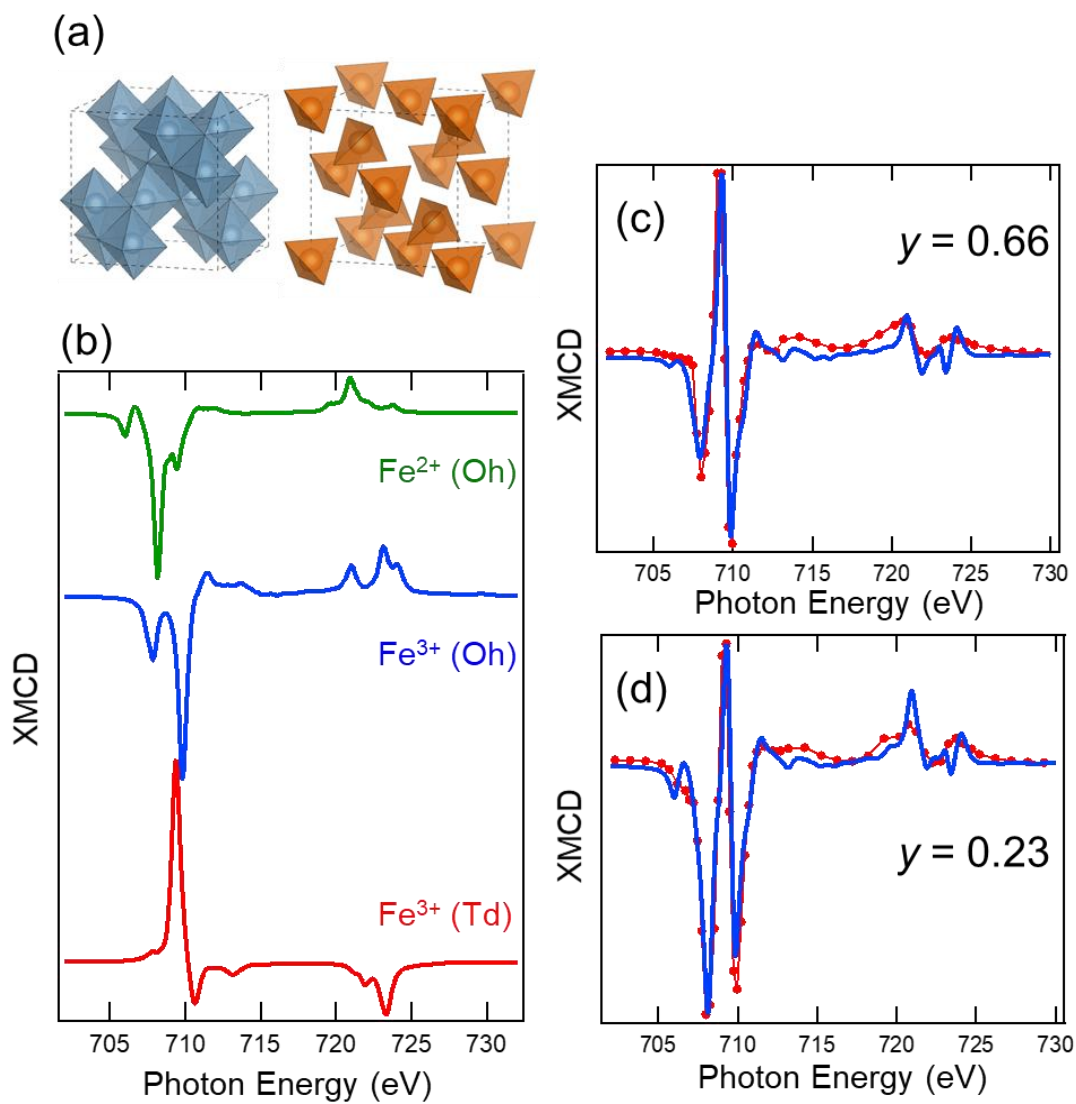

Fig. 7 Ligand-field multiplet cluster-model calculation of $Fe^{2+}$ and $Fe^{3+}$ states. (a) Fragments of $O_h$ and $T_d$ symmetries used in the calculation drawn by VESTA (K. Momma and F. Izumi, 2011). (b) Calculated XMCD spectra of $Fe^{3+}$ ($O_h$, $T_d$) and $Fe^{2+}$ ($O_h$). (c,d) XMCD spectra of experimental (dots) and LFM calculation (lines) for $y = 0.23$ and 0.66.

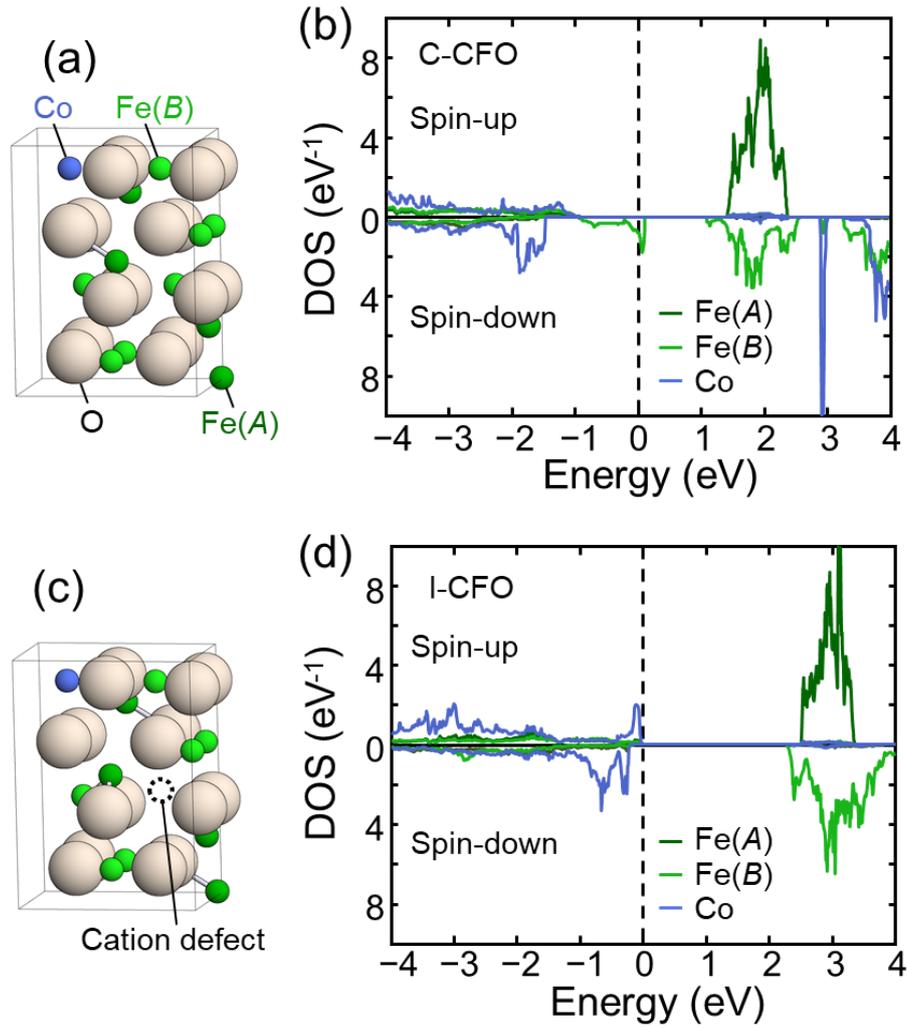

Fig. 8 (a) Crystal structure of the Co$_1$Fe$_{11}$O$_{16}$ (C-CFO) used in first principles calculation. The white, dark green, light green and blue spheres are O, Fe (*A*-site), Fe (*B*-site) and Co atoms, respectively. (b) The local density of states for Co$_1$Fe$_{11}$O$_{16}$ of spin up and spin down states. The Fermi energy was set to 0 eV. (c) Crystal structure of the Co$_1$Fe$_{10}$O$_{16}$ (I-CFO) used in first principles calculation. The white, dark green, light green and blue spheres are O, Fe (*A*-site), Fe (*B*-site) and Co atoms, respectively. The dotted circle represents a cation defect. (d) The local density of states for Co$_1$Fe$_{10}$O$_{16}$ spin up and spin down states.